\documentstyle{article}
\newtheorem{rem}{Remark}
\newtheorem{theor}{Theorem}

\begin{document}
\begin{center}
\Large{\bf 10-DIM EINSTEIN SPACES MADE UP ON \\[1mm]BASIS OF 6-DIM
RICCI- FLAT SPACES \\[2mm] AND 4-DIM EINSTEIN
SPACES}\vspace{4mm}\normalsize
\end{center}
 \begin{center}
\Large{\bf Valery Dryuma}\vspace{4mm}\normalsize
\end{center}
\begin{center}
{\bf Institute of Mathematics and Informatics AS Moldova, Kishinev}\vspace{4mm}\normalsize
\end{center}
\begin{center}
{\bf  E-mail: valery@dryuma.com;\quad cainar@mail.md}\vspace{4mm}\normalsize
\end{center}
\begin{center}
{\bf  Abstract}\vspace{4mm}\normalsize
\end{center}

   Some examples of ten-dimensional vacuum Einstein spaces ($^{10}R_{ij}=0$)
  made up on basis of four-dimensional Ricci-flat ($^{4}R_{ij}=0$) Einstein spaces and
    six-dimensional Ricci-flat spaces ($^{6}R_{ij}=0$) defined by solutions of the classical
$Sin-Gordon$ equation are constructed.

    The properties of  geodesics for such type of the spaces are discussed.

\section{Introduction}

   The properties of classical four dimensional  Einstein spaces are dependent on the
   energy-momentum tensor of  matter $T_{ik}$
\begin{equation} \label{dryuma:eq1}
R_{i j}=\frac{8
\pi\kappa}{c^4}\left(T_{ik}-\frac{1}{2}g_{ik}T\right).
\end{equation}

    Tensor $T_{ik}$ is self-dependent object in the Einstein theory of gravitation and in general
    does not has geometric description.

    The most popular approach to the geometric description of the
    matter tensor $T_{ik}$ and their relation  with a Ricci-flat tensor $^{4}R_{ij}$ of Space-Time takes
    place within the bounds of the Kaluza-Klein theories using the string theory on Calaby-Yau manifolds.

    At the same time both substance - Space-Time and the Matter are considered as a single
    whole.

    We present here a new possibilities for such type of considerations.

\section{Three-dimensional space of constant negative curvature}

   We start from a three-dimensional space endowed with the metrics in form
\begin{equation}\label{dryuma:eq2}
 ds^2=dx^2+2\cos(u(x,y)dxdy+dy^2+A(x,y)^2dz^2.
\end{equation}

   The condition on the space to be the space of constant
   negative curvature
\[
R_{i j k l}-\lambda\left(g_{i k}g_{j l}-g_{i l}g_{j k}\right)=0
\]
in the case $\lambda=-1$ lead to the compatible system of
equations
\begin{equation} \label{dryuma:eq3}
{\frac {\partial ^{2}}{\partial x\partial y}}u(x,y)-\sin(u(x,
y))=0,
\]
\[
 {\frac {\partial ^{2}}{\partial x\partial y}}A(x,y)-\,A(x,y)
\cos(u(x,y))=0,
\]
\[
{\frac {\partial ^{2}}{\partial {x}^{2}}}A(x,y)-{\frac
{\cos(u(x,y)) \left ({\frac {\partial }{\partial x}}u(x,y)\right
){\frac {\partial } {\partial
x}}A(x,y)}{\sin(u(x,y))}}-\,A(x,y)+{\frac {\left ({ \frac
{\partial }{\partial x}}u(x,y)\right ){\frac {\partial }{
\partial y}}A(x,y)}{\sin(u(x,y))}}=0,
\]
\[
{\frac {\partial ^{2}}{\partial {y}^{2}}}A(x,y)-{\frac
{\cos(u(x,y)) \left ({\frac {\partial }{\partial y}}u(x,y)\right
){\frac {\partial } {\partial
y}}A(x,y)}{\sin(u(x,y))}}-\,A(x,y)+{\frac {\left ({ \frac
{\partial }{\partial x}}A(x,y)\right ){\frac {\partial }{
\partial y}}u(x,y)}{\sin(u(x,y))}}=0.
\end{equation}

     As it follows that for any solution $u(x,y)$ of the $Sin-Gordon$ equation  one possible to find
         the function $A(x,y)$ by solving corresponding linear system
    of equations.

    So the following theorem is valid  (\cite{dryuma:dr1}).
    \begin{theor}
    Three-dimensional spaces having the metric ~(\ref{dryuma:eq2})
    with the functions $A(x,y)$ and $u(x,y)$ defined by a given
    system of equations are the spaces
    of constant negative curvature $\lambda=-1$ .
\end{theor}

    To take one example.

    The simplest solution of the equation

\[
{\frac {\partial ^{2}}{\partial x\partial y}}u(x,y)-\sin(u(x,
y))=0
\]
is given by
\[
u(x,y)=4\arctan(\exp(x+y)).
\]

    At this condition the linear system looks as
\[
{\frac {\partial ^{2}}{\partial x\partial y}}A(x,y)+{\frac {\left
(6\, {e^{2\,x+2\,y}}-1-{e^{4\,x+4\,y}}\right )A(x,y)}{\left
(1+{e^{2\,x+2\, y}}\right )^{2}}}=0,
\]
\[
\left (1\!-\!{e^{4\,x+4\,y}}\right )A(x,y)\!+\!\left
(-2\,{e^{2\,x+2\,y}}-{e^{ 4\,x+4\,y}}-1\right ){\frac {\partial
}{\partial y}}A(x,y)\!+\!\left (-6\,
{e^{2\,x+2\,y}}\!+\!1\!+\!{e^{4\,x+4\,y}}\right ){\frac {\partial
}{\partial x }}A(x,y)+\]\[+\left (-1+{e^{4\,x+4\,y}}\right ){\frac
{\partial ^{2}}{
\partial {x}^{2}}}A(x,y)=0,
\]
\[
\left (1\!-\!{e^{4\,x+4\,y}}\right )A(x,y)\!+\!\left
(-6\,{e^{2\,x+2\,y}}+1+{e ^{4\,x+4\,y}}\right ){\frac {\partial
}{\partial y}}A(x,y)+\left (-2\,
{e^{2\,x+2\,y}}-{e^{4\,x+4\,y}}-1\right ){\frac {\partial
}{\partial x }}A(x,y)+\]\[+\left (-1+{e^{4\,x+4\,y}}\right ){\frac
{\partial ^{2}}{
\partial {y}^{2}}}A(x,y)=0
\]

    The simplest solution of this system is
\[
A(x,y)={\frac {{e^{x+y}}}{1+{e^{2\,x+2\,y}}}}.
\]

   In result we get the example of the metric of constant negative
   curvature
\begin{equation} \label{dryuma:eq4}
ds^2=dx^2+2\cos(4\arctan(\exp(x+y)))dx
dy+dy^2+\left(\frac{\exp(x+y)}{1+\exp(2x+2y)}\right)^2 dz^2.
\end{equation}

\begin{rem}

   In 3-dimensional geometry the density of  Chern-Simons invariant defined by
\begin{equation} \label{dryuma:eq5}
CS(\Gamma)=\epsilon^{i j k}(\Gamma^p_{i q}\Gamma^q_{k
p;j}+\frac{2}{3}\Gamma^p_{i q}\Gamma^q_{j r}\Gamma^r_{k p})
\end{equation}
has an important role.

   For the metric (\ref{dryuma:eq2}) one get
\[
CS(\Gamma)=0.
\]

\end{rem}

    Our construction of ten dimensional Einstein space be composed from a few steps.

    The first one is the creation of the six dimensional basic space with a
    necessary properties.

    With this aim we use the notion of the Riemann extension of a given three-dimensional
    space.

\section{Six dimensional Riemann extensions of the metrics\\[1mm] of constant negative curvature}

   The Riemann extension of riemannian or nonriemannian spaces can be constructed with the
   help of the Chistoffel coefficients $\Gamma^i_{jk}$ of corresponding Riemann space or
   with the help of connection coefficients $\Pi^i_{jk}$ of affinely connected space.

   The metrics of the Riemann extension of a given n-dimensional riemannian space looks as
\begin{equation} \label{dryuma:eq6}
^{2n}ds^2=-2\Gamma^i_{jk}dx^jdx^k \psi_i+2dx^id\psi_i,
\end{equation}
where $\psi_i$ are some additional coordinates.

    To give an examples of the properties of Riemann
    extensions.

    \begin{theor}
    The Riemann extension of 3-dimensional  space of constant curvature is a six-dimensional
     symmetrical space
\[
^{6}R_{i j k l;m}=0.
\]
\end{theor}

\begin{theor}
    The Riemann extension of Ricci-flat $^{n}R_{i k}=0$ space is a Ricci-flat $^{2n}R_{i k}=0$
    space.
\end{theor}
 \begin{theor}
    The  spaces with conditions
\[
R_{i j; k}+R_{k i;j}+R_{j k;i}=0.
\]
on the Ricci-tensor conserve such conditions after the Riemann
extension.
\end{theor}

   After the Riemann extension of three-dimensional space one get a six-dimensional
     space having the signature $[+++---]$

    In case of the metric (\ref{dryuma:eq2}) we have a following components of the
    Christoffel symbols
\[
\Gamma^2_{33}={\frac {A(x,y)\cos(u(x,y)){\frac {\partial
}{\partial x}}A(x,y)}{ \left (\sin(u(x,y))\right )^{2}}}-{\frac
{A(x,y){\frac {\partial }{
\partial y}}A(x,y)}{\left (\sin(u(x,y))\right )^{2}}},
\]\[\Gamma^1_{33}=-{\frac {A(x,y){\frac {\partial }{\partial
x}}A(x,y)}{\left (\sin(u(x, y))\right )^{2}}}+{\frac
{A(x,y)\cos(u(x,y)){\frac {\partial }{
\partial y}}A(x,y)}{\left (\sin(u(x,y))\right )^{2}}}
,\]\[ \Gamma^3_{23}={\frac {{\frac {\partial }{\partial
y}}A(x,y)}{A(x,y)}} ,\quad\Gamma^2_{22}={\frac {\cos(u(x,y)){\frac
{\partial }{\partial y}}u(x,y)}{\sin(u(x,y) )}} ,\quad
\Gamma^1_{22}=-{\frac {{\frac {\partial }{\partial
y}}u(x,y)}{\sin(u(x,y))}}
 ,\]\[\Gamma^3_{13}={\frac {{\frac {\partial
}{\partial x}}A(x,y)}{A(x,y)}} ,\quad\Gamma^2_{11}=-{\frac {{\frac
{\partial }{\partial x}}u(x,y)}{\sin(u(x,y))}} ,\quad
\Gamma^1_{11}={\frac {\cos(u(x,y)){\frac {\partial }{\partial
x}}u(x,y)}{\sin(u(x,y) )}}.
\]

  So using these values  we get from the ~(\ref{dryuma:eq6}) the metrics of
six-dimensional extension
\begin{equation}\label{dryuma:eq7}
^{6}ds^2=\left (2\,{\frac {\left ({\frac {\partial }{\partial x}
}u(x,y)\right )V}{\sin(u(x,y))}}-2\,{\frac {\cos(u(x,y))\left
({\frac {\partial }{\partial x}}u(x,y)\right
)U}{\sin(u(x,y))}}\right ){{\it dx}}^{2}-4\,{\frac {\left ({\frac
{\partial }{\partial x}}A(x,y) \right )W{\it dx}\,{\it
dz}}{A(x,y)}}+\]\[+2\,{\it dx}\,{\it dU}+\left (-2 \,{\frac
{\cos(u(x,y))\left ({\frac {\partial }{\partial y}}u(x,y) \right
)V}{\sin(u(x,y))}}+2\,{\frac {\left ({\frac {\partial }{
\partial y}}u(x,y)\right )U}{\sin(u(x,y))}}\right ){{\it dy}}^{2}-4\,{
\frac {\left ({\frac {\partial }{\partial y}}A(x,y)\right )W{\it
dy}\, {\it dz}}{A(x,y)}}+2\,{\it dy}\,{\it dV}+\]\[+A(x,y)\left
(\!-\!2\,{\frac {V \cos(u(x,y)){\frac {\partial }{\partial
x}}A(x,y)}{\left (\sin(u(x,y)) \right )^{2}}}\!+\!2\,{\frac
{V{\frac {\partial }{\partial y}}A(x,y) }{\left
(\sin(u(x,y))\right )^{2}}}\!+\!2\,{\frac {U{\frac {
\partial }{\partial x}}A(x,y)}{\left (\sin(u(x,y))\right )^{2}}}\!-\!2\,{
\frac {U\cos(u(x,y)){\frac {\partial }{\partial y}}A(x,y)}{ \left
(\sin(u(x,y))\right )^{2}}}\right ){{\it dz}}^{2}\!+\]\[+2\,{\it
dz}\,{ \it dW},
\end{equation}
where $U,V,W$ are the set of additional coordinates.

   Remark that we have the relations between the set of solutions of the system (\ref{dryuma:eq3})
    and corresponding six-dimensional spaces endowed with the metrics
    (\ref{dryuma:eq7}).
In particular we get the set of such type
    of the spaces corresponded the soliton solutions of the
    $Sin-Gordon$-equation.

    Ricci-tensor of the metric (\ref{dryuma:eq7}) has non zero
    components $^{6}R_{ik}\neq 0$ and
    the next problem is transformation of the space (\ref{dryuma:eq7}) into the Ricci flat $^{6}R_{i j}=0$
    space defined by the solutions of the system (\ref{dryuma:eq3}).

\section{Six-dimensional Ricci-flat space}

    For construction of the Ricci-flat six-dimensional space we
    consider the metrics which are  conformal to the metrics
    (\ref{dryuma:eq7})
\begin{equation}  \label{dryuma:eq8}
 ^{6}d \tilde s^2=\frac{^{6}ds^2}{A(x,y)^2},
\end{equation}
with the function $A(x,y)$ defined by the solutions of the system
(\ref{dryuma:eq3}).

   The calculation with a GRTensorII (\cite{dryuma:dr12}) show that the Ricci-tensor of the metrics
(\ref{dryuma:eq8})  has only one non zero component
 \[
 R_{zz}\neq 0.
 \]

In explicit form it looks as
\begin{equation} \label{dryuma:eq9}
R_{zz}=-\left (\cos(u(x,y))\right )^{2}\left (A(x,y)\right
)^{2}+2\,\cos(u (x,y))\left ({\frac {\partial }{\partial
y}}A(x,y)\right ){\frac {
\partial }{\partial x}}A(x,y)+\]\[+\left (A(x,y)\right )^{2}-\left ({
\frac {\partial }{\partial x}}A(x,y)\right )^{2}-\left ({\frac
{\partial }{\partial y}}A(x,y)\right )^{2}
 \end{equation}

    It is interested to note that this quantity is an integral of the system
    ~(\ref{dryuma:eq3}).

    In fact after differentiation the quantity (\ref{dryuma:eq9}) on the variables $x$
    or $y$ we get the expressions containing the second order
    derivatives  of the functions $A_{xx}$, $A_{xy}$, $A_{yy}$. The substitution
    of the corresponding values from the system (\ref{dryuma:eq3}) make these
expressions  vanish.

\begin{rem}

   The meaning of the value $R_{zz}$ is dependent from the choice of the solutions of the system
    (\ref{dryuma:eq3}).

      For example  on the  solution

\[
u(x,y)=4\arctan(\exp(x+y)),\quad
A(x,y)=\frac{\exp(x+y)}{1+\exp(2x+2y)}
\]
we get
\[
R_{zz}=0.
\]

At the same time on the singular  solution of the system
(\ref{dryuma:eq3})
\[
u(x,y)=4\,\arctan({e^{i\left (x-y\right )}}),\quad A(x,y)={\frac
{{e^{i\left(x-y\right)}}}{1+{e^{2\,i\left(x-y\right)}}}}
\]
we get
\[
R_{zz}=-4\,{\frac {{e^{2\,i\left (3\,y+x\right )}}-2\,{e^{4\,i\left (y+x
\right )}}+{e^{2\,i\left (y+3\,x\right )}}}{\left
({e^{2\,iy}}+{e^{2\, ix}}\right )^{4}}}.
\]
\end{rem}

      So we get the family of six-dimensional metrics (\ref{dryuma:eq8}) depending from the solutions of
      the $Sin-Gordon$-equation and having only one non zero component of the Ricci-tensor.

      Now for achievement of our aim we must add some additional terms into the expressions for the metrics
       (\ref{dryuma:eq8}) in such a way that
      the Ricci-flat six-dimensional manifold defined by the solutions of
      the $Sin-Gordon$-equation will be obtained.

      Remark that there are a lot possibilities to make that.

      All of them
      are connected with the freedom in choice of connections
      coefficients at the construction of our six-dimensional manifold
      in result of the extensions of concrete three-dimensional space( there are only eight components
       of the connections in the expression (\ref{dryuma:eq8})).

      For example adding the term
\[
\frac{F(x,y)}{A(x,y)^2}U
\]
into the component $g_{zz}$ of
      metric tensor $g_{ij}$ give rise to the metric
\begin{equation} \label{dryuma:eq10}
{{\it ^{6}ds}}^{2}=\left (2\,{\frac {\left ({\frac {\partial
}{\partial x} }u(x,y)\right )V}{\left (A(x,y)\right
)^{2}\sin(u(x,y))}}-2\,{\frac { \cos(u(x,y))\left ({\frac
{\partial }{\partial x}}u(x,y)\right )U}{ \left (A(x,y)\right
)^{2}\sin(u(x,y))}}\right ){{\it dx}}^{2}-4\,{ \frac {\left
({\frac {\partial }{\partial x}}A(x,y)\right )W{\it dx}\, {\it
dz}}{\left (A(x,y)\right )^{3}}}+\]\[+2\,{\frac {{\it dx}\,{\it
dU}}{ \left (A(x,y)\right )^{2}}}+\left (-2\,{\frac
{\cos(u(x,y))\left ({ \frac {\partial }{\partial y}}u(x,y)\right
)V}{\left (A(x,y)\right )^{ 2}\sin(u(x,y))}}+2\,{\frac {\left
({\frac {\partial }{\partial y}}u(x, y)\right )U}{\left
(A(x,y)\right )^{2}\sin(u(x,y))}}\right ){{\it dy}} ^{2}-4\,{\frac
{\left ({\frac {\partial }{\partial y}}A(x,y)\right )W{ \it
dy}\,{\it dz}}{\left (A(x,y)\right )^{3}}}+\]\[+2\,{\frac {{\it
dy}\,{ \it dV}}{\left (A(x,y)\right )^{2}}}+\]\[+\left (-2\,{\frac
{\cos(u(x,y))V{ \frac {\partial }{\partial x}}A(x,y)}{A(x,y)\left
(\sin(u(x,y))\right )^{2}}}+2\,{\frac {V{\frac {\partial
}{\partial y}}A(x,y)}{A(x,y) \left (\sin(u(x,y))\right
)^{2}}}+{\frac {F(x,y)U}{\left (A(x,y) \right )^{2}}}+2\,{\frac
{U{\frac {\partial }{\partial x}}A(x,y)}{A(x, y)\left
(\sin(u(x,y))\right )^{2}}}\right ){{\it dz}}^{2}-\]\[\left
(-2\,{\frac {\cos(u(x,y))U{\frac {
\partial }{\partial y}}A(x,y)}{A(x,y)\left (\sin(u(x,y))\right )^{2}}}
\right ){{\it dz}}^{2}+2\,{\frac {{\it dz}\,{\it dW}}{\left
(A(x,y) \right )^{2}}}.
\end{equation}

     This metric is a Ricci-flat
\[
^{6}R_{ij}=0
\]
 but non a flat! $R_{ijkl}\neq 0$
     if the function $F(x,y)$ is satisfied the equation
\begin{equation} \label{dryuma:eq11}
{\frac {\partial }{\partial x}}F(x,y)-{\frac {-\left ({\frac {
\partial }{\partial x}}u(x,y)\right )A(x,y)\cos(u(x,y))+3\,\left ({
\frac {\partial }{\partial x}}A(x,y)\right
)\sin(u(x,y))}{A(x,y)\sin(u (x,y))}}+\]\[+4\,{\frac {\left
(A(x,y)\right )^{2}\left (\cos(u(x,y)) \right
)^{2}-2\,\cos(u(x,y))\left ({\frac {\partial }{\partial y}}A(x,
y)\right ){\frac {\partial }{\partial x}}A(x,y)-\left
(A(x,y)\right )^ {2}+\left ({\frac {\partial }{\partial
y}}A(x,y)\right )^{2}+\left ({ \frac {\partial }{\partial
x}}A(x,y)\right )^{2}}{-1+\left (\cos(u(x,y ))\right
)^{2}}}=\]\[=0.
\end{equation}

     In result we get the set of Ricci-flat six-dimensional
     manifolds
     defined by the solutions of the $Sin-Gordon$-equation.

     According  the geometric approach such type of manifold may be suitable arena for the string theory.

\section{$3\times 2$+4=10}

   Now we present the construction of ten-dimensional Ricci-flat
   space
\[
^{10}R_{ik}=0
\]
 made up on six-dimensional Ricci-flat manifold and
four-dimensional Ricci-flat Einstein space.

    By way as example of four-dim Einstein space will be considered the Schwarzschild
    Space-Time with the metric
\[
ds^2=(1-M/r)c^2dt^2-r^2(\sin(\theta)^2d\phi^2+d\theta^2)-dr^2/(1-M/r).
\]

   As the  six-dimensional Ricci-flat space will be used the space with the
metric (\ref{dryuma:eq8}) with a suitable additional term.
\begin{equation} \label{dryuma:eq12}
{{\it ^{10}ds}}^{2}=\left (2\,{\frac {\left ({\frac {\partial
}{\partial x} }u(x,y)\right )V}{\left (A(x,y)\right
)^{2}\sin(u(x,y))}}-2\,{\frac { \cos(u(x,y))\left ({\frac
{\partial }{\partial x}}u(x,y)\right )U}{ \left (A(x,y)\right
)^{2}\sin(u(x,y))}}\right ){{\it dx}}^{2}-4\,{ \frac {\left
({\frac {\partial }{\partial x}}A(x,y)\right )W{\it dx}\, {\it
dz}}{\left (A(x,y)\right )^{3}}}+2\,{\frac {{\it dx}\,{\it dU}}{
\left (A(x,y)\right )^{2}}}+\]\[+\left (-2\,{\frac
{\cos(u(x,y))\left ({ \frac {\partial }{\partial y}}u(x,y)\right
)V}{\left (A(x,y)\right )^{ 2}\sin(u(x,y))}}+2\,{\frac {\left
({\frac {\partial }{\partial y}}u(x, y)\right )U}{\left
(A(x,y)\right )^{2}\sin(u(x,y))}}\right ){{\it dy}} ^{2}-4\,{\frac
{\left ({\frac {\partial }{\partial y}}A(x,y)\right )W{ \it
dy}\,{\it dz}}{\left (A(x,y)\right )^{3}}}+2\,{\frac {{\it dy}\,{
\it dV}}{\left (A(x,y)\right )^{2}}}+\]\[+\left (\!-\!2\,{\frac
{\cos(u(x,y))V{ \frac {\partial }{\partial x}}A(x,y)}{A(x,y)\left
(\sin(u(x,y))\right )^{2}}}\!+\!2\,{\frac {V{\frac {\partial
}{\partial y}}A(x,y)}{A(x,y) \left (\sin(u(x,y))\right
)^{2}}}+2\,{\frac {U{\frac {\partial }{\partial x}}A(x,y)}{A(x,
y)\left (\sin(u(x,y))\right )^{2}}}\!-\!2\,{\frac
{\cos(u(x,y))U{\frac {
\partial }{\partial y}}A(x,y)}{A(x,y)\left (\sin(u(x,y))\right )^{2}}}
\right ){{\it dz}}^{2}+\]\[+2\,{\frac {{\it dz}\,{\it dW}}{\left
(A(x,y) \right )^{2}}}-{{\it dr}}^{2}\left (1-{\frac {M}{r}}\right
)^{-1}-{r}^ {2}\left (\sin(\theta)\right )^{2}\left (d(\phi)\right
)^{2}-{r}^{2} \left (d(\theta)\right )^{2}+\left ({c}^{2}-{\frac
{{c}^{2}M}{r}} \right ){{\it dt}}^{2}+\]\[+{\frac {F(x,y)U}{\left
(A(x,y) \right )^{2}}}dz^2.
\end{equation}

 In result we have the metric of the union space which is composed from two
 independent spaces.

    In particular the geodesic equations of full 10D-space are
    dissolved on two independent subsystems of equations - for the coordinates $(x,y,z,U,V,W)$
    and $(r,\theta,\phi,t)$.

    Under such condition on the metric the influence of 6D-space which is the medium of the Matter ( analogue of
    Calaby-Yau space!) on the properties of 4D-Space-Time is absent.

    For description of interaction between both structures it is necessary to introduce
    additional terms into the expression for the metric.

    As example we consider a following metric of 10D-space
\begin{equation} \label{dryuma:eq13}
^{10}ds^2=\left (2\,{\frac {\left ({\frac {\partial }{\partial
x}}u(x,y)\right ) V}{\left (A(x,y)\right
)^{2}\sin(u(x,y))}}-2\,{\frac {\cos(u(x,y)) \left ({\frac
{\partial }{\partial x}}u(x,y)\right )U}{\left (A(x,y) \right
)^{2}\sin(u(x,y))}}\right ){{\it dx}}^{2}-4\,{\frac {\left ({
\frac {\partial }{\partial x}}A(x,y)\right )W{\it dx}\,{\it dz}}{
\left (A(x,y)\right )^{3}}}+2\,{\frac {{\it dx}\,{\it dU}}{\left
(A(x, y)\right )^{2}}}+\]\[+\left (-2\,{\frac {\cos(u(x,y))\left
({\frac {
\partial }{\partial y}}u(x,y)\right )V}{\left (A(x,y)\right )^{2}\sin(
u(x,y))}}+2\,{\frac {\left ({\frac {\partial }{\partial y}}u(x,y)
\right )U}{\left (A(x,y)\right )^{2}\sin(u(x,y))}}\right ){{\it
dy}}^{ 2}-4\,{\frac {\left ({\frac {\partial }{\partial
y}}A(x,y)\right )W{ \it dy}\,{\it dz}}{\left (A(x,y)\right
)^{3}}}+2\,{\frac {{\it dy}\,{ \it dV}}{\left (A(x,y)\right
)^{2}}}+\]\[+\left (\!-\!2\,{\frac {\cos(u(x,y))V{ \frac {\partial
}{\partial x}}A(x,y)}{A(x,y)\left (\sin(u(x,y))\right
)^{2}}}\!+\!2\,{\frac {V{\frac {\partial }{\partial
y}}A(x,y)}{A(x,y) \left (\sin(u(x,y))\right )^{2}}}\!+\!2\,{\frac
{U{\frac {\partial }{
\partial x}}A(x,y)}{A(x,y)\left (\sin(u(x,y))\right )^{2}}}\!-\!2\,{\frac
{\cos(u(x,y))U{\frac {\partial }{\partial y}}A(x,y)}{A(x,y)\left
(\sin (u(x,y))\right )^{2}}}\right ){{\it dz}}^{2}+\]\[+2\,{\frac
{{\it dz}\,dW}{ \left (A(x,y)\right )^{2}}}-{{\it dr}}^{2}\left
(1-{\frac {M}{r}} \right )^{-1}-{r}^{2}\left (\sin(\theta)\right
)^{2}\left (d(\phi) \right )^{2}-{r}^{2}\left (d(\theta)\right
)^{2}+\left ({c}^{2}-{ \frac {{c}^{2}M}{r}}\right ){{\it
dt}}^{2}+\]\[+{\frac {H(r,t)}{ \left (A(x,y)\right )^{2}}}+{\frac
{F(x,y)U}{\left (A(x,y)\right )^{2} }},
\end{equation}
 where  the new additional term ${\frac {H(r,t)}{ \left (A(x,y)\right
)^{2}}}$  was added into the expression for the metric
(\ref{dryuma:eq12}).

     Taking into consideration
      the relation (\ref{dryuma:eq11}) the conditions on the metric (\ref{dryuma:eq13}) to be a Ricci-flat
      lead to the following equation  for determination of the function $H(r,t)$
\begin{equation} \label{dryuma:eq14}
   -{r}^{3}{\frac {\partial ^{2}}{\partial {t}^{2}}}H(r,t)+r{c}^{2}\left
(r-M\right )^{2}{\frac {\partial ^{2}}{\partial
{r}^{2}}}H(r,t)+{c}^{2 }\left (2\,r-M\right )\left (r-M\right
){\frac {\partial }{\partial r} }H(r,t)=0.
 \end{equation}
\begin{rem}
The solutions of the equation (\ref{dryuma:eq14}) can be presented
in form \[H(r,t)=F_2(t)F_1(r)
\]
where the functions $F_2(t)$ and $F_1(r)$ satisfy the equations
\[
{\frac {d^{2}}{d{t}^{2}}}{\it F_2}(t)={\it \_c}_{{1}}{\it
F_2}(t){c} ^{2},
\]
 and
\[{\frac {d^{2}}{d{r}^{2}}}{\it F_1}(r)={\frac {{r}^{2}{\it F_1}(r){
\it \_c}_{{1}}}{\left (r-M\right )^{2}}}-{\frac {\left ({\frac
{d}{dr} }{\it F_1}(r)\right )\left (2\,r-M\right )}{\left
(r-M\right )r}}.
\]
where ${\it \_c}_{{1}}$ is arbitrary parameter.

The first equation is elementary and the second equation is more
complicated  and equivalent the equation without the first
derivative
\[{\frac {d^{2}}{d{r}^{2}}}{\it F_3}(r)-1/4\,{\frac {{\it F_3}(r)
\left (4\,{r}^{4}{\it \_c}_{{1}}-{M}^{2}\right )}{{r}^{2}\left
({r}-{M}\right )^2}}=0
\]
where
\[
{\it F_1}(r)={\frac {{\it F_3}(r)}{\sqrt {\left (r-M\right )r}}}.
\]
\end{rem}

    So the metric of ten-dimensional Ricci-flat spaces dependent from the solutions of
     $Sin-Gordon$-equation  has been constructed and our problem is solved.

     Let us discuss the properties of geodesic equations in this
     case.

\section{On geodesic equations}

    For simplicity's sake we consider a following  equations for geodesic of the metric
    (\ref{dryuma:eq13})
\[
{\frac {d^{2}}{d{s}^{2}}}r(s)=\left (r-M\right )\left
(\sin(\theta) \right )^{2}\left ({\frac {d}{ds}}\phi(s)\right
)^{2}+\left (r-M \right )\left ({\frac {d}{ds}}\theta(s)\right
)^{2}-1/2\,{\frac { \left (r-M\right ){c}^{2}M\left ({\frac
{d}{ds}}t(s)\right )^{2}}{{r}^ {3}}}+\]\[+1/2\,{\frac {M\left
({\frac {d}{ds}}r(s)\right )^{2}}{\left (r-M \right
)r}}-1/2\,{\frac {\left (r-M\right )\left ({\frac {
\partial }{\partial r}}H(r,t)\right )\left ({\frac {d}{ds}}z(s)\right
)^{2}}{r\left (A(x,y)\right )^{2}}},
\]
and
\[{\frac {d^{2}}{d{s}^{2}}}t(s)=-{\frac {M\left ({\frac {d}{ds}}r(s)
\right ){\frac {d}{ds}}t(s)}{\left (r-M\right )r}}+1/2\,{\frac
{r\left ({\frac {\partial }{\partial t}}H(r,t)\right )\left
({\frac {d} {ds}}z(s)\right )^{2}}{\left (r-M\right ){c}^{2}\left
(A(x,y)\right )^ {2}}}.
\]

    Comparison of these expressions with standard equations for
    geodesics of the  Schwarzshild Space-Time show us essential
    distinctions between them.

    In fact we observe that the expressions for geodesics of general ten-dimensional Ricci-flat space are
    contained additional terms depending from the function $A(x,y)$ which in turn is dependent from
solutions of the $Sin-Gordon$-equation  arising  in context of the
theory of the spaces of constant negative curvature and
corresponding Ricci-flat six-dimensional space (see
(\ref{dryuma:eq3})).

    From a given point of view the reason of appearance  of new terms in geodesic equations is pure geometric and
    is motivated by consideration of the problem in the spirit of the Kaluza-Klein
    theories.

    At the same time six-dimensional Ricci-flat space defined
    by solutions of the $Sin-Gordon$ equation
    stands in the role of Calaby -Yau variety.

     This fact offer the new challenge for the problem of geometrical description of joint properties
     of Space-Time and Matter.
\begin{rem}
     In the articles of author (\cite{dryuma:dr9}-\cite{dryuma:dr10}) analogous
     approache to the problem of description of joint properties of
      Space-Time and Matter was considered on the basis of the
      Korteveg-de Vrize (KdV) equations (Cylindrical KdV, mKdV) which are described some classes  of three -
      dimensional metrics of zero curvature (\cite{dryuma:dr1}).

      Six-dimensional Ricci-flat Riemann  extensions of 3-dimensional metrics depending from solutions of
      KdF-equations were constructed.

      The properties of geodesics of ten-dimensional Ricci-flat spaces
      combined  on basis of interactive   six-dimensional Ricci-flat  spaces
      and Schwarzschild or  E.Kasner Space-Time were investigated.
\end{rem}

 \end{document}